\documentclass[pra,amsmath,amssymb,twocolumn]{revtex4}
\usepackage{mathrsfs}
\usepackage{amssymb}

\input epsf.tex
\usepackage{graphicx}
\usepackage{amsthm}

\begin{document}

\title{Semi-device independent random number expansion without entanglement}

\author{Hong-Wei Li$^{1,2}$, Zhen-Qiang Yin$^{1a}$, Yu-Chun Wu$^{1}$, Xu-Bo Zou$^{1}$,  Shuang Wang$^{1}$, Wei Chen$^{1}$,  Guang-Can Guo$^1$,  Zheng-Fu Han$^{1b}$ }

 \affiliation
 {$^1$ Key Laboratory of Quantum Information,University of Science and Technology of China,Hefei, 230026,
 China\\$^2$ Zhengzhou Information Science and Technology Institute, Zhengzhou, 450004,
 China}

 \date{\today}
\begin{abstract}
By testing the classical correlation violation between two systems,
the random number can be expanded and certified without applying
classical statistical method. In this work, we propose a new random
number expansion protocol without entanglement, and the randomness
can be guaranteed only by the 2-dimension quantum witness violation.
Furthermore, we only assume that the dimensionality of the system
used in the protocol has a tight bound, and the whole protocol can
be regarded as a semi-device independent black-box scenario.
Comparing with the device independent random number expansion
protocol based on entanglement, our protocol is much easier to
implement and test.
\end{abstract}
\maketitle

\section{ Introduction}\label{Introduction}

True random numbers have significant applications in numerical
simulation, lottery games, biological systems and cryptography
\cite{RANDOM}. More particularly, security of the quantum key
distribution (QKD) protocol \cite{BB84} is based on the random
selection of the state preparation and measurement, if the state
preparation and measurement are known by the eavesdropper precisely,
she can apply the man-in-the-middle attack \cite{Man} to get all of
the secret information without being discovered. True random numbers
should be unpredictable for the third party, so most of true random
number generation protocols are based on unpredictable physical
processes
\cite{RANDOM1,RANDOM2,RANDOM3,RANDOM4,RANDOM5,RANDOM6,RANDOM7,RANDOM8}.
Unfortunately, the true random number generated by these protocols
can only be characterized with the classical statistical method,
such as the Statistical Test Suite from NIST et al. \cite{NIST
1,NIST 2}. Inspired by the device independent quantum information
processing based on non-local correlations of entanglement particles
\cite{DI 1,DI 2,DI 3}, Colbeck et al. \cite{Colbeck 1,Colbeck 2}
have proposed the true random number generation protocol based on
the GHZ test, Pironio et al. \cite{Pironio 1,Pironio 2} have
proposed the true random number generation protocol certified by the
Bell inequality violation, they also have given a proof of concept
experimental demonstration of their protocol by approximately one
metre distance. The true random number generation protocol based on
entanglement, which require no assumption about the internal working
of the device both in two states measurement sides, thus the true
random number can not be generated with only the classical method,
and the randomness of their experimental result can only be
certified by the Bell inequality violation. Since the protocol
require the pre-established true random number to select the
measurement basis, it also can be called device independent random
number expansion protocol correspondingly. Comparing with the random
number generation protocol certified by the classical statistical
method, the device independent random number expansion protocol
offers a new method to quantify unequivocally the observed random
numbers.

Both of the two device independent random number expansion protocols
strongly suggest that only entanglement based protocols are suitable
for establishing the quantified true random numbers \cite{Colbeck
1,Colbeck 2,Pironio 1,Pironio 2}. However, the entanglement based
protocol has much more complicated experimental setups comparing
with the one-way system, where the first black box prepares an
arbitrary quantum state, sends it to the other black box for
performing an arbitrary measurement. Furthermore, most commerical
true random number generation systems are based on one-way
protocols. Inspired by the method of device-independent test of the
classical and quantum dimensions given by Gallego et al.
\cite{Witness 1}, Pawlowski et al. \cite{Witness 2} have proposed a
semi-device independent one-way QKD protocol with 4 input states and
2 measurement bases, security of which was based on the
2-dimensional quantum witness and the quantum random access code.
Here, we propose the one-way semi-device independent random number
expansion protocol without entanglement, the randomness of which can
be quantified with 2-dimension quantum witness, and the experimental
demonstration can be established by combining the commerical QKD
setup with different modulation protocols, the randomness of which
can be proved in the following section by applying the numerical
calculation method. Similar to Colbeck and Pironio's models, our
protocol require no assumption about the internal working of the
state preparation and measurement device, except that the
2-dimensional quantum system and collective attacks are bounded.
However, we need the quantum state to be prepared and measured in
the same safe area, the quantum state and classical information
should not be divulged to the eavesdropper in the unsafe area.

\section {Model description}
In this section, we give the semi-device independent random number
expansion protocol, where only two black boxes should be considered.
The two black boxes can be used for illustrating the state
preparation and measurement respectively, detailed scenario can be
depicted precisely in Fig. 1.

\begin{figure}[!h]\center
\resizebox{7cm}{!}{
\includegraphics{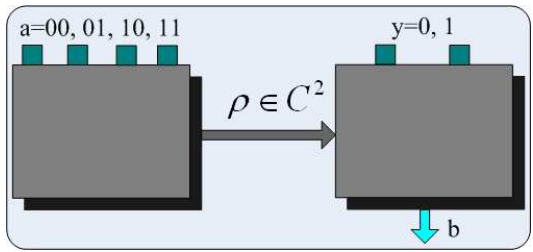}}
\caption{Semi-device independent random number expansion protocol.
The protocol require the state preparation black box and the state
measurement black box respectively, both of the two black boxes are
in the same safe area.}
\end{figure}

In the semi-device independent random number expansion protocol, we
randomly select four classical input bits $a\in\{00,01,10,11\}$ in
the first black box, when pressing the button $a$, the first black
box will emit the classical or quantum state $\rho_{a}$, then the
prepared state $\rho_{a}$ will be sent to the second black box
correspondingly. When pressing the button $y=\{0,1\}$, the second
black box will emit the measurement outcome $b=\{0,1\}$. We suppose
that only the 2-dimensional system will be considered in this
protocol, that is $\rho_{a}\in C^2$.

Formally, we can estimate the probability distribution by repeating
this procedure many times, which can be illustrated precisely as the
following equation,
\begin{equation}
\begin{array}{lll}
P(b|ay)=tr(\rho_{a}M_y^b),
\end{array}
\end{equation}
where, $M_y^b$ is the measurement operator acting on two dimension
Hilbert space with the input parameter $y$ and the output parameter
$b$ by considering the prepared state $\rho_{a}$. In this protocol,
the true random number can be produced by only considering the date
table $P(b|ay)$. More precisely, we do not require any assumption on
how the probability was obtained with two black boxes, except that
the state preparation and measurement can be guaranteed with
2-dimension quantum witness.

We will use the following expectation value to illustrate the
probability distribution for the convenient analysis in the
following section,
\begin{equation}
\begin{array}{lll}
E_{ay}=P(b=0|ay),\\
P(b=0|ay)+P(b=1|ay)=1,
\end{array}
\end{equation}
where the set of probability distributions $E_{ay}$ can be used for
illustrating the quantum dimension witness. In the theoretical side,
two types of 2-dimension quantum witness have been proposed
\cite{Witness 1,Witness 2}, we will apply the following tight
2-dimension classical witness in our security analysis
\begin{equation}
\begin{array}{lll}                 \label{TTT}
T\equiv
E_{000}+E_{001}+E_{010}-E_{011}\\~~~~~-E_{100}+E_{101}-E_{110}-E_{111}\leq2,
\end{array}
\end{equation}
where we only consider the 4 state preparation and 2 measurement
basis case in this inequation. The other similar expression with 3
state preparation and 2 measurement bases case has also been given
in Ref. \cite{Witness 1}, but the 2-dimension quantum witness in
this case can not be used for expanding the true random number.

More precisely, the tight 2-dimension quantum witness can be given
as the following inequation (More detailed information about this
inequation can also be found in Ref. \cite{Witness 2})
\begin{equation}
\begin{array}{lll}                 \label{TT2}
T\equiv
E_{000}+E_{001}+E_{010}-E_{011}\\~~~~~-E_{100}+E_{101}-E_{110}-E_{111}\leq
2.828,
\end{array}
\end{equation}
the maximal value of the two-dimension quantum witness can be
calculated numerically. More interestingly, it also can be analyzed
by applying the 2-to-1 quantum random access code protocol
\cite{Witness 2,RAC}, where Alice receives two uniformly distributed
bits and sends the encoded physical system to Bob, Bob is asked to
guess one of Alice's bits randomly. This 2-dimension quantum witness
is the main tool to analyze the proposed random number expansion
protocol, and our main result is to establish the relationship
between the randomness of the measurement outcome and its expected
2-dimension quantum witness violation.

We quantify the randomness of the measurement outcome $b$
conditioned on the input values $a,y$ by the min-entropy
\cite{renner}
\begin{equation}
\begin{array}{lll}
H_{\infty}(B|A,Y)\equiv-log_{2}[max_{b,a,y}P(b|a,y)].
\end{array}
\end{equation}
From this equation, we can find that the purpose of this paper is to
obtain the upper bound of the conditional probability distribution
$P(b|a,y)$ for a given 2-dimension quantum witness $T$. More
precisely, the maximal probability distribution
$max_{b,a,y}P(b|a,y)$ denotes the solution to the following
optimization problem:
\begin{equation}
\begin{array}{lll}
max_{b,a,y}P(b|a,y)\\
subject~~to~:~\\
E_{000}+E_{001}+E_{010}-E_{011}
-E_{100}+E_{101}-E_{110}-E_{111}=T\\
E_{ay}=tr(\rho_{a}M_y^0)
\end{array}
\end{equation}
where the optimization is carried arbitrary quantum states
$\{\rho_{00},\rho_{01},\rho_{10},\rho_{11}\}$ and measurement
operators $\{M_0^0,M_1^0\}$ defined over 2-dimension Hilbert space.
In the most general case, we should consider the positive operator
valued measure (POVM) $\{M_0^0,M_0^1\}$ and $\{M_1^0,M_1^1\}$, where
 $M_0^0+M_0^1=M_1^0+M_1^1=I$.
Fortunately, Masanes \cite{Masanes} has proved that only the
projective measurement should be considered in case of 2-observable
and 2-measurement outcomes has been considered. Since $T$ is the
linear expression of the probabilities, we can only consider pure
states \cite{Witness 1} preparation in our numerical calculation.
Without loss of generality, the state preparation and measurement in
our numerical calculation can be illustrated precisely with the
following equations respectively,

\begin{equation}
\begin{array}{lll}
\rho_{a}=|\varphi(a)\rangle\langle\varphi(a)|,
\end{array}
\end{equation}

\begin{equation}
\begin{array}{lll}
|\varphi(a)\rangle= \left(\begin{array}{ccc}
cos (\frac{\theta_{a}}{2})\\
e^{i\eta_{a}}sin (\frac{\theta_{a}}{2})
\end{array}\right),
\end{array}
\end{equation}

\begin{equation}
\begin{array}{lll}
M_0^0= \left(\begin{array}{ccc}
1&0\\
0&0 \end{array}\right),
\end{array}
\end{equation}

\begin{equation}
\begin{array}{lll}
M_1^0= \left(\begin{array}{ccc}
cos^2(\frac{\theta}{2})&\frac{1}{2}e^{-i\eta} sin(\theta) \\
\frac{1}{2}e^{i\eta} sin(\theta)&sin^2(\frac{\theta}{2})
\end{array}\right),
\end{array}
\end{equation}
where $a\in\{00,01,10,11\}$, $0\leq\theta_{a},\theta\leq\pi$,
$0\leq\eta_{a},\eta\leq2\pi$. By applying the maximization problem,
we get the min-entropy bound of the measurement outcome for given
2-dimension quantum witness $T$, detailed expression of the
relationship between the 2-dimension quantum witness and the
min-entropy bound can be depicted precisely in Fig. 2.
\begin{figure}[!h]\center
\resizebox{7.5cm}{!}{
\includegraphics{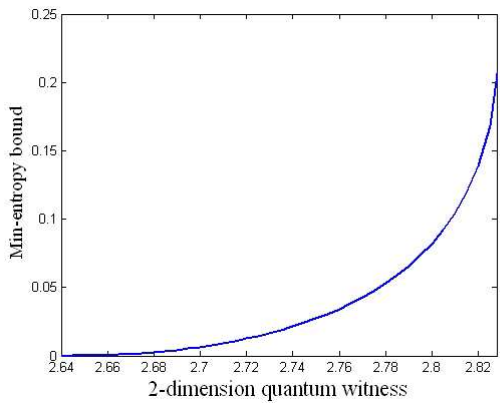}}
\caption{The relationship between the 2-dimension quantum witness
and the the min-entropy bound. The min-entropy starts at zero in the
2-dimension classical witness case, systems that violate the
2-dimension quantum witness 2.64 have a positive min-entropy.}
\end{figure}

The calculation result shows that if the the violation of the
2-dimension quantum witness is larger than 2.64, the semi-device
independent true random number can be expanded correspondingly. The
maximal value of the min-entropy bound in our numerical calculation
is 0.206, which can be satisfied in case of the 2-dimension quantum
 witness violation is 2.828.

\section {Example description}

In this section, we give a particular protocol to illustrate the
smei-device independent random number expansion protocol. This
protocol is equal to the (2,1,0.85) quantum random access code
protocol \cite{RAC1,RAC2,RAC,Witness 2}. In this particular
protocol, the state preparation in the first black box can be
illustrated precisely as the following equations
\begin{equation}
\begin{array}{lll}
|\varphi(00)\rangle=cos(\frac{\pi}{8})|0\rangle+sin(\frac{\pi}{8})|1\rangle,\\
|\varphi(01)\rangle=cos(\frac{7\pi}{8})|0\rangle+sin(\frac{7\pi}{8})|1\rangle,\\
|\varphi(10)\rangle=cos(\frac{3\pi}{8})|0\rangle+sin(\frac{3\pi}{8})|1\rangle,\\
|\varphi(11)\rangle=cos(\frac{5\pi}{8})|0\rangle+sin(\frac{5\pi}{8})|1\rangle.
\end{array}
\end{equation}
For the state measurement in the second black box, we will apply the
two projective measurements with the following bases
\begin{equation}
\begin{array}{lll}
\{M_0^0=|0\rangle\langle0|,~~~~M_0^1=|1\rangle\langle1|\},\\
\{M_1^0=|+\rangle\langle+|,~~M_1^1=|-\rangle\langle-|\}.\\
\end{array}
\end{equation}
where, $|+\rangle=\frac{1}{\sqrt{2}}(|0\rangle+|1\rangle),
|-\rangle=-\frac{1}{\sqrt{2}}(|0\rangle-|1\rangle)$. The 2-dimension
quantum witness in this protocol is 2.828, which is the maximal
2-dimension quantum witness violation. Combining this state
preparation and measurement protocol with the true random number
extraction analysis result, we can numerically calculate min-entropy
bound of the expanded random bit is 0.206. Note that we only need
true random numbers $a$ and $y$ to estimate dimension witness value,
no more random numbers should be pre-established by two black boxes,
thus our random number expansion protocol only need few random
number seed.

Since the BB84 protocol is also based on the 4 input states and 2
measurement bases case, one natural question is considering whether
the BB84 protocol can be used for generating the true random number
in our randomness analysis. Unfortunately, it can not be used for
generating the true random number because of it does not violate the
2-dimension quantum witness 2.64. More precisely, the state
preparation in the BB84 protocol can be illustrated as
\begin{equation}
\begin{array}{lll}
|\widetilde{\varphi}(00)\rangle=|0\rangle,|\widetilde{\varphi}(01)\rangle=|-\rangle,\\|\widetilde{\varphi}(10)\rangle=|+\rangle,|\widetilde{\varphi}(11)\rangle=|1\rangle,
\end{array}
\end{equation}
the measurement bases are equal to the (2,1,0.85) random access code
case. Then the dimension witness achieves $T=2$, which indicates
that no true random number can be generated by considering the
semi-device independent random number expansion protocol.

\section {discussion}

We have proposed a new true random number expansion protocol in this
paper, we can quantify the randomness with 2-dimension quantum
witness violation, not based on the classical statistical method.

Comparing with the quantified random number expansion protocol based
on entanglement, we give a much simpler method, and our protocol
does not need any entanglement. Unfortunately, since the maximal
ratio of the expanded random number is 0.206, our protocol has a
much lower random number expansion efficiency. However, since our
semi-device independent protocol is much easier to implement than
the full device independent protocol based on entanglement, thus the
semi independent protocol will generate much more random numbers
than the full device independent protocol in the same period of
time. It is an open question to discuss if people can find a much
higher efficiency random number expansion protocol in the future
research with the similar method. Similar to the security analysis
given by Pironio et al. \cite{Pironio 2}, it also will be very
interesting to analytically prove the min-entropy bound by
considering the quantum dimension witness violation.

Device-independent quantum information has attracted much attentions
for its higher level security comparing with the protocol based on
trusted devices. Combining the semi-device independent random number
expansion protocol with the device independent QKD protocol, we hope
to get a much higher security than the QKD protocol based solely on
some mathematical methods certified random numbers.

\section {acknowledgement}
The author Hong-Wei Li would like to thank Dr. Marcin Pawlowski for
his helpful discussion and comments. This work was supported by the
National Basic Research Program of China (Grants No. 2011CBA00200
 and No. 2011CB921200), National Natural Science Foundation of China (Grant NO. 60921091, 10974193), and China Post
doctoral Science Foundation (Grant No. 20100480695).
 To whom correspondence should be addressed, Email:
$^a$yinzheqi@ustc.edu.cn, $^b$zfhan@ustc.edu.cn.

\end{document}